\documentclass[12pt]{iopart}

\usepackage{iopams}
\usepackage{xcolor}
\usepackage{amsthm}
\usepackage{amsfonts}
\usepackage{graphicx}
\usepackage{subfigure}
\usepackage{epstopdf}
\usepackage{ulem}
\usepackage{soul}
\usepackage{amstext}
\usepackage{amssymb} 
\usepackage{latexsym}
\usepackage[english]{babel}

\begin{document}

\title{Synthetic gauge fields in synthetic dimensions: Interactions and chiral edge modes}

\author{Simone Barbarino$^{1}$, Luca Taddia$^{2,3}$, Davide Rossini$^{1}$,\\
Leonardo Mazza$^{1}$, and Rosario Fazio$^{4,1}$}

\address{$^{1}$NEST, Scuola Normale Superiore \& Istituto Nanoscienze-CNR, I-56126 Pisa, Italy}
\address{$^{2}$Scuola Normale Superiore, I-56126 Pisa, Italy}
\address{$^{3}$CNR - Istituto Nazionale di Ottica, UOS di Firenze LENS, I-50019 Sesto Fiorentino, Italy}
\address{$^{4}$The Abdus Salam International Centre for Theoretical Physics (ICTP), I-34151 Trieste, Italy}

\begin{abstract}
Synthetic ladders realized with one-dimensional alkaline-earth(-like) fermionic gases 
and subject to a gauge field represent a promising environment for the investigation 
of quantum Hall physics with ultracold atoms. 
Using density-matrix renormalization group calculations, we study how the quantum 
Hall-like chiral edge currents  are affected by repulsive atom-atom interactions. 
We relate the properties of such currents to the asymmetry of the spin resolved momentum 
distribution function, a quantity which is easily addressable in state-of-art experiments. 
We show that repulsive interactions significantly stabilize the quantum Hall-like
helical region and enhance the chiral currents. 
Our numerical simulations are performed for atoms with two and three internal spin states.
\end{abstract}

\section{Introduction}

One of the most noticeable hallmarks of topological insulators is the presence of 
robust {\it gapless edge modes}~\cite{topins}. 
Their first experimental observation goes back to the discovery of the quantum Hall 
effect~\cite{qhe}, where the existence of chiral edge states is responsible for the striking 
transport properties of the Hall bars. 
The physics of edge states has recently peeked out also in the arena 
of ultracold gases~\cite{Atala_2014, Mancini_2015, Stuhl_2015}, triggered by the new exciting developments in the implementation of topological models and synthetic gauge potentials for neutral cold atoms~\cite{Dalibard_2011, Struck_2012, Hauke_2012, Goldman_2013, Goldman_2014}.

Synthetic gauge potentials in cold atomic systems have already led to the experimental study of  
Bose-Einstein condensates coupled to a magnetic field~\cite{Lin_2009} or with an effective
spin-orbit coupling~\cite{Lin_2011}, and more recently to lattice models 
with non-zero Chern numbers~\cite{Aidelsburger_2013, Miyake_2013, Jotzu_2014, Aidelsburger_2014}
and frustrated ladders~\cite{Atala_2014}. 
In a cold-gas experiment, the transverse dimension of a two-dimensional setup
does not need to be a \textit{physical} dimension, i.e. a dimension in real space: 
an extra \textit{synthetic} dimension on a given \textit{d}-dimensional lattice 
can be engineered taking advantage of the internal atomic degrees of freedom~\cite{Boada_2012}. 
The crucial requirement is that each of them has to be coupled to two other states 
in a sequential way through, for example, proper Raman transitions induced by two laser beams. 
In this situation, it is even possible to generate gauge fields in synthetic lattices~\cite{Celi_2014}. 

In this work we focus on one-dimensional systems with a finite synthetic dimension 
coupled to a synthetic gauge field, i.e. \textit{frustrated ladders}.
The study of such ladders traces back to more than thirty years ago, 
when frustration and commensurate-incommensurate transitions have been addressed 
in Josephson networks~\cite{kardar1,kardar2}. 
Thanks to the experimental advances with optical lattices, these systems are now reviving 
a boost of activity. Both bosonic (see, e.g., Refs.~\cite{dhar, petrescu, grudsdt, piraud, tokuno}) 
and fermionic (see, e.g., Refs.~\cite{roux, sun, Barbarino_2015, Zeng_2015, Cornfeld_2015, Mazza_2015, Budich_2015, Lacki_2015}) 
systems have been considered. 
The emerging phenomenology is amazingly rich, ranging from new phases with chiral 
order~\cite{dhar} to vortex phases~\cite{piraud} or fractional Hall-like phases 
in fermionic systems~\cite{Barbarino_2015, Cornfeld_2015}, just to give some examples. 
Very recently, two experimental groups~\cite{Mancini_2015, Stuhl_2015} have observed persistent 
spin currents in one dimensional gases of $^{173}$Yb (fermions) and $^{87}$Rb (bosons) 
determined by the presence of such gauge field.
Within the framework of the synthetic dimension, such 
\textit{helical} spin currents can be regarded as the \textit{chiral} edge states 
of a two-dimensional system and are reminiscent of the edge modes of the Hall effect. 

Up to now, the study of edge currents in optical lattices has mainly focused on aspects 
related to the single-particle physics and a systematic investigation 
of the interaction effects is missing. 
{Repulsive} interactions considerably affect the properties of the edge modes: 
this is well known in condensed matter, where the fractional quantum Hall regime~\cite{fqhe} 
can be reached for proper particle fillings and for sufficiently strong Coulomb interactions. 
In view of the new aforementioned experiments in bosonic~\cite{Stuhl_2015} 
and fermionic~\cite{Mancini_2015} atomic gases, a deeper 
understanding of the role of repulsive interactions in these setups 
is of the uttermost importance.

Here we model the experiment on the frustrated $n$-leg ladder performed in 
Ref.~\cite{Mancini_2015} and analyze, by means of density-matrix renormalization group (DMRG) 
simulations, how atom-atom repulsive interactions modify the edge physics of the system
(in this article we disregard the effects of an harmonic confinement and of the temperature).
We concentrate on the momentum distribution function, which has been used 
in the experiment to indirectly probe the existence of the edge currents. 
The purpose of this article is twofold. First, we want to present numerical evidence 
that helical modes, reminiscent of the chiral currents of the integer quantum Hall effect, 
can be stabilized by repulsive interactions. Second, we want to discuss the influence 
of interactions on experimentally measurable quantities that witness the chirality of the modes. 
In this context the words ``chiral'' and ``helical'' can be interchanged, 
depending whether one considers a truly one-dimensional system with an internal degree 
of freedom or a synthetic ladder.
There is an additional important point to be stressed when dealing with synthetic ladders 
in the presence of interactions. The many-body physics of alkaline-earth(-like) 
atoms (like Ytterbium) with nuclear spin $I$ larger than $1/2$ is characterized 
by a SU($2I+1$) symmetry~\cite{Gorshkov_2010, Cazalilla_2014}. 
When they are viewed as ($2I+1$)-leg ladders, the interaction is strongly anisotropic, 
i.e. it is short-range in the physical dimension and long-range in the synthetic dimension. 
This situation is remarkably different from {the typical} condensed-matter
systems and may lead to quantitative differences especially when considering narrow ladders,
as in Ref.~\cite{Mancini_2015}.

The paper is organized as follows. In the next section we introduce the model describing 
a one-dimensional gas of earth-alkaline(-like) atoms with nuclear spin $I\geq 1/2$. 
In order to make a clear connection with the experiment of Ref.~\cite{Mancini_2015}, 
we briefly explain how this system can be viewed as a ($2I+1$)-leg ladder. 
Moreover, we present a discussion of the single-particle spectrum to understand 
the main properties of the edge currents in the non-interacting regime and to identify 
the regimes where the effects of repulsive interactions are most prominent. 
Then, in Sec.~\ref{obs-sec} we introduce two quantities, evaluated by means of the DMRG 
algorithm, that characterize the edge currents: the (spin-resolved) momentum 
distribution function and the average current derived from it. 
In Sec.~\ref{results} we present and comment our results; we conclude with a summary in Sec.~\ref{conclusions}.

\section{Synthetic gauge fields in synthetic dimensions}
\label{model}

\subsection{The model}

\begin{figure}
  \centering
  \includegraphics[width=\linewidth]{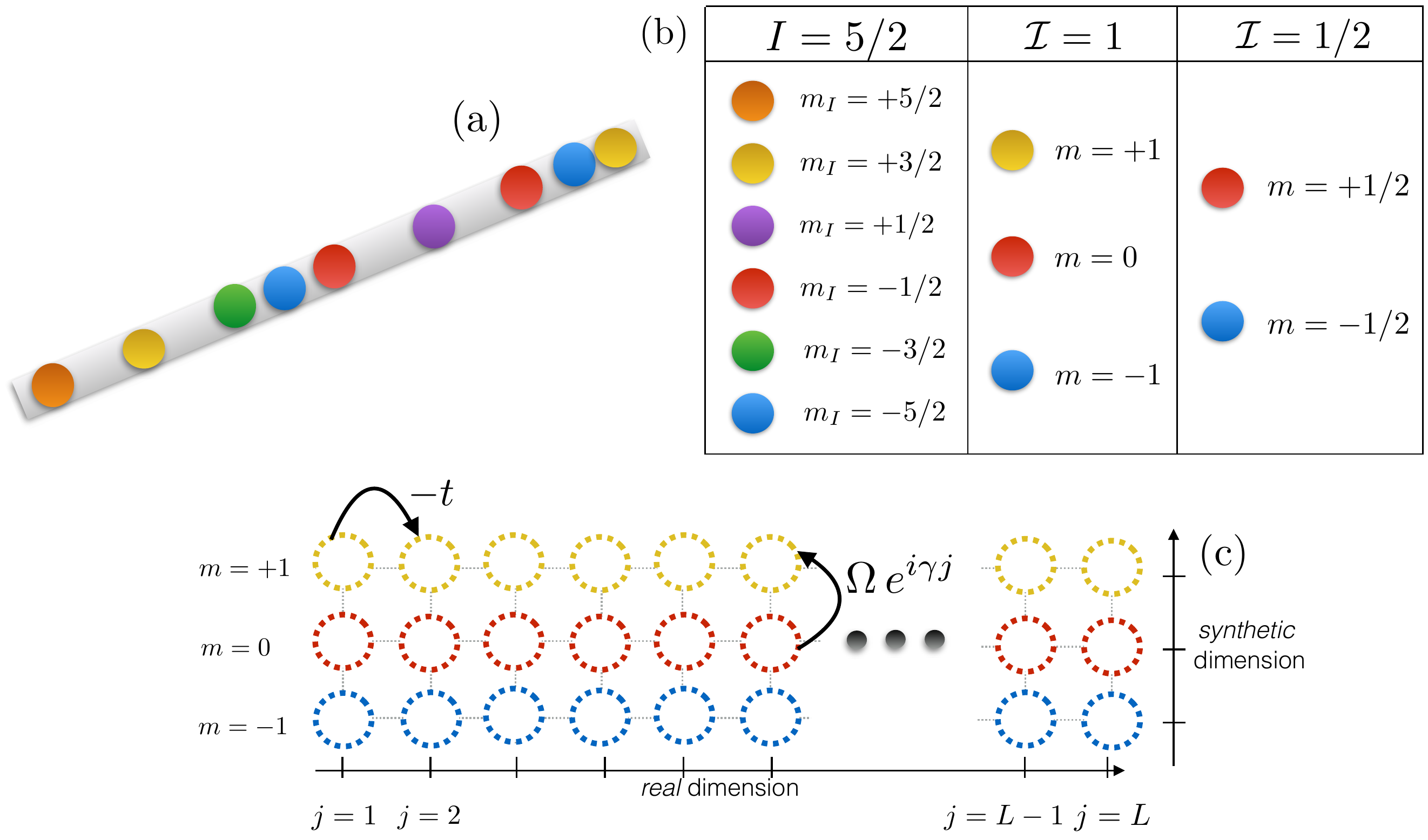}
  \caption{Implementation of $\hat{H} = \hat{H}_0 + \hat{H}_1$ in a cold-atom system.
    (a) Sketch of a one-dimensional atomic gas with nuclear spin $I=5/2$, e.g. $^{173}$Yb.
    (b) Definition of the effective spins $\mathcal I=1$ and $\mathcal I=1/2$
    as in the experimental implementation with $^{173}$Yb of Ref.~\cite{Mancini_2015}. 
    (c) Graphical representation of the non-interacting Hamiltonian 
    in the synthetic-dimension picture, for the case ${\mathcal I}=1$.
    }
  \label{ladder}
\end{figure}

We consider a one-dimensional gas of fermionic earth-alkaline-(like) neutral atoms 
characterized by a large and tunable nuclear spin $I$, see Fig.~\ref{ladder}(a).
Based on the predictions of Ref.~\cite{Gorshkov_2010}, Pagano {\it et al.}
have experimentally showed that, by conveniently choosing the populations 
of the nuclear-spin states, the number of atomic species can be reduced at will 
to $2\mathcal{I}+1$, giving rise to an effective atomic spin $\mathcal{I}\leq I$~\cite{Pagano_2014}. 
We stress that $I$ has to be an half-integer to enforce the fermionic statistics, 
while $\mathcal{I}$ can also be an integer, see Fig.~\ref{ladder}(b). 
Moreover, as extensively discussed in Refs.~\cite{Boada_2012, Celi_2014}, 
the system under consideration can be both viewed as a mere one-dimensional gas 
with $2\mathcal{I}+1$ spin states or as a ($2\mathcal{I}+1$)-leg ladder, see Fig.~\ref{ladder}(c). 

When loaded into an optical lattice, the Hamiltonian can be written as~\cite{Gorshkov_2010}:
\begin{equation}
  \hat{H}_0 =- t \sum_j \sum_{m=-\mathcal{I}}^\mathcal{I}  \left( \hat{c}^\dagger_{j, m} \hat{c}_{j+1,  m} + 
  \mathrm{H.c.} \right) + U \sum_j \sum_{m < m'} \hat{n}_{j,m} \hat{n}_{j,m'} \,, \label{Hubbard}
\end{equation}
where $\hat{c}_{j,m}$ ($\hat{c}^\dagger_{j,m}$) annihilates (creates) a spin-$m$ fermion 
($m=-\mathcal{I}, \ldots, \mathcal{I}$) at site $j=1,\ldots,L$ and 
$\hat{n}_{j,m}=\hat{c}^\dagger_{j,m} \hat{c}_{j,m}$; 
$t$ is the hopping amplitude, while $U$ is the strength of the 
SU($2\mathcal{I} + 1$)-invariant interaction; the first sum in the hopping term runs 
over $j=1, \ldots, L-1$ if open boundary conditions (OBC) in the real dimension are considered,
or over $j=1, \ldots, L$ if periodic boundaries (PBC) are assumed. 
Hereafter we set $\hbar=1$.
The Hamiltonian~(\ref{Hubbard}), also known as the SU($2\mathcal{I}+1$) 
Hubbard model, has attracted considerable attention in the last few decades, 
see e.g. Refs.~\cite{Assaraf_1999, Szirmai_2005, Buchta_2007, Manmana_2011}.

The presence of two additional laser beams can induce a coupling between spin-states 
with $\Delta m = \pm 1$ of amplitude $\Omega_m$ endowed with a running 
complex phase factor $e^{i\gamma j}$. 
For simplicity, in the following we assume that $\Omega_m$ does not depend on $m$ 
and set $\Omega_m=\Omega$. 
The coupling $\Omega$ is related to the amplitude of the laser beams, 
while the phase $\gamma$ {depends on} their wavelength and relative propagation angle. 
Explicitly, the Hamiltonian gets a contribution of the form
\begin{equation}
  \hat{H}_1 = \sum_j \sum_{m=-\mathcal{I}}^{\mathcal{I}-1} \; \Omega_m \left( 
  e^{-i\gamma j} \hat{c}^\dagger_{j, m} \hat{c}_{j, m+1}+ \mathrm{H.c.} \right) \, .
\end{equation}

As already mentioned, the system characterized by the Hamiltonian $\hat{H} \equiv \hat{H}_0+\hat{H}_1$ is equivalent to a $(2\mathcal{I}+1)$-leg ladder 
{where the} coordinate in the transverse direction is given by the effective-spin 
index $m=-\mathcal{I}, \dots, \mathcal{I}$. 
For all purposes, such direction can be regarded as a synthetic dimension 
with sharp edges; in this framework, the Hamiltonian $\hat{{H}}_1$ describes 
the hopping in the synthetic dimension and introduces
a constant magnetic field perpendicular to the ladder with dimensionless
magnetic flux $+\gamma$ per plaquette.
The peculiarity of our synthetic ladder resides in the interaction term,  
which is $SU(2\mathcal{I}+1)$ invariant: it therefore describes an on-site 
interaction in the real dimension and a long-range interaction in the synthetic one. 

Since the Hamiltonian $\hat H$ is not translationally invariant, 
for later convenience, we perform the unitary transformation 
$\hat{d}_{j,m}=\hat{\mathcal{U}}\hat{c}_{j,m}\hat{\mathcal{U}}^\dagger=e^{-im \gamma j }\hat{c}_{j,m}$ 
such that $\hat{\mathcal{U}}(\hat{H}_0+\hat{H}_1)\hat{\mathcal{U}}^\dagger = \hat{\mathcal{H}}_0+\hat{\mathcal{H}}_1 = \hat{\mathcal{H}}$ reads
\begin{eqnarray}
  \hat{\mathcal{H}}_0 & = & -t \sum_j \sum_{m=-\mathcal{I}}^\mathcal{I} 
  \left(e^{i\gamma m} \hat{d}^\dagger_{j, m} \hat{d}_{j+1,  m} + \mathrm{H.c.} \right) + 
  U \sum_j \sum_{m < m'} \hat{\nu}_{j,m} \hat{\nu}_{j,m'} \,,
  \\
  \hat{\mathcal{H}}_1 & = & \sum_j \sum_{m=-\mathcal{I}}^{\mathcal{I}-1} 
  \left( \Omega_m \; \hat{d}^\dagger_{j, m} \hat{d}_{j, m+1} + \mathrm{H.c.} \right) \,,
\end{eqnarray}
where $\hat{\nu}_{j,m}= \hat{d}^\dagger_{j,m} \hat{d}_{j,m}$. 
Assuming PBC in the real dimension, the quadratic part of $\hat{\mathcal{H}}$ 
can be diagonalized in Fourier space, in terms of the operators 
$\hat{{d}}_{p,m}=L^{-1/2} \sum_{j=1}^L e^{ik_pj}  \hat{d}_{j,m}$, 
with $k_p={2\pi p}/{L}$ and $p \in \{-L/2, \ldots, L/2-1\}$.

\subsection{Non-interacting helical liquid}

\begin{figure}[p]
  \centering
  \includegraphics[width=\linewidth, trim=0cm -0.3cm  0cm 0cm, clip=true]{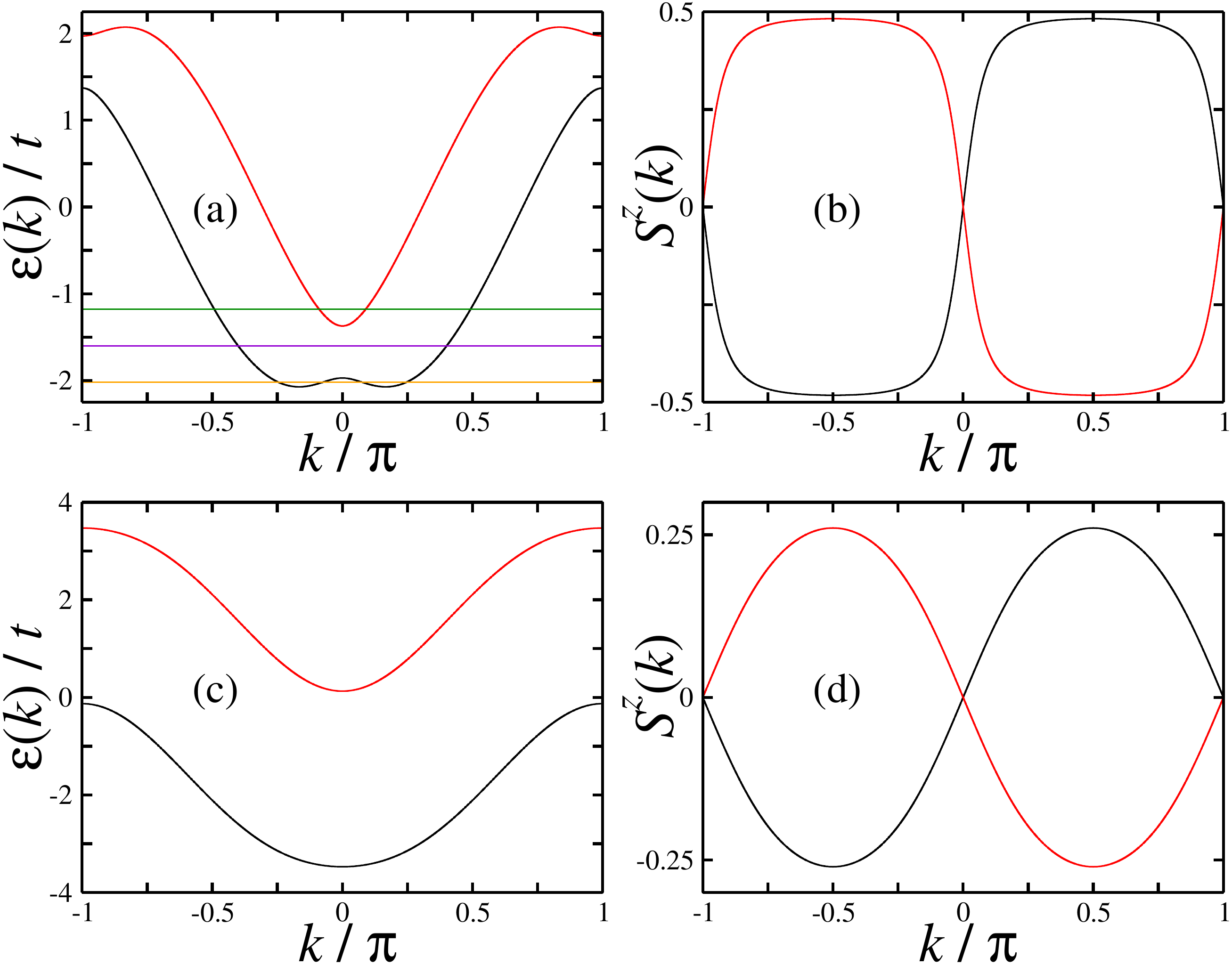}
  \includegraphics[width=\linewidth]{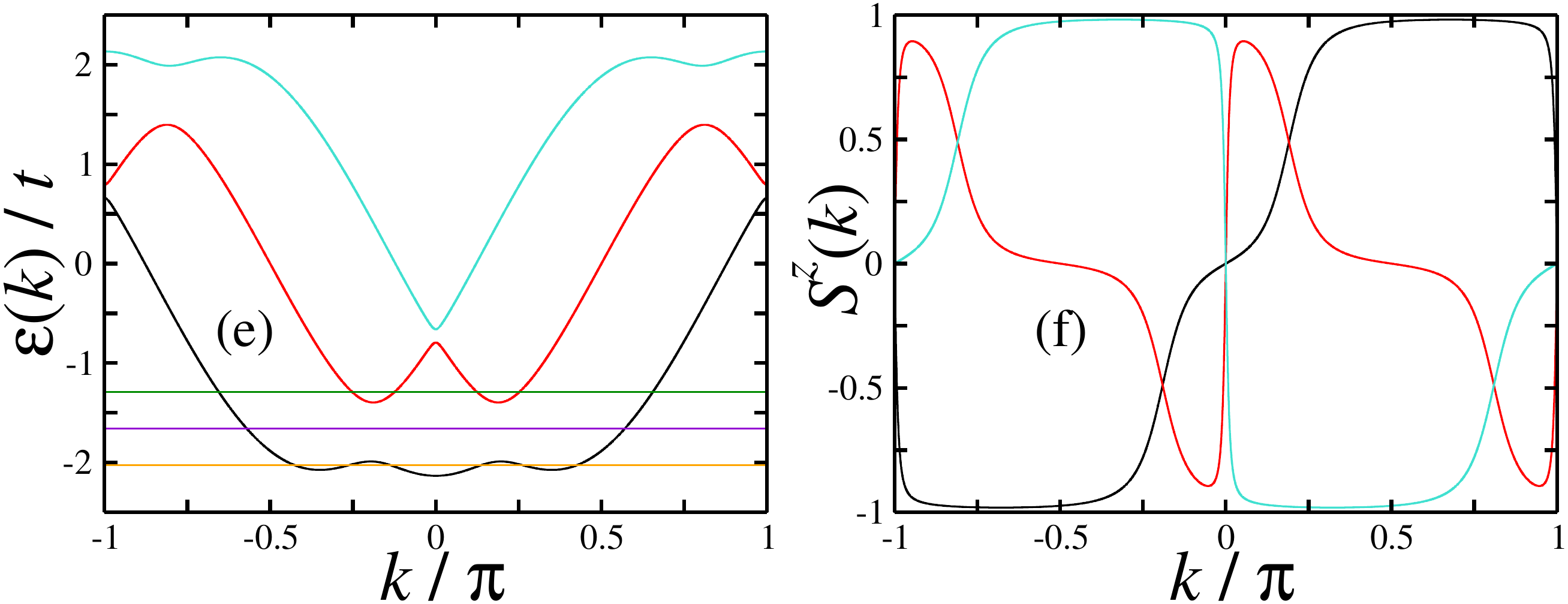}
  \caption{Spectral properties of $\hat{\mathcal{H}}$ in the non-interacting case. 
    Left panels: energy spectra; right panels: spin polarization along the $z$ axis
    of the quasi-momentum single-particle eigenstates for several cases
    (lines with the same colors are in correspondence). 
    Panels (a)-(b): $\mathcal{I}=1/2$ and WRC regime ($\Omega/t=0.3$).
    Panels (c)-(d): $\mathcal{I}=1/2$ and SRC regime ($\Omega/t=1.8$).
    Panels (e)-(f): $\mathcal{I}=1$ and WRC regime ($\Omega/t=0.1$).
    In all the situations, we assumed $\gamma=0.37\pi$, PBC and $L\rightarrow\infty$. 
    In panels (a) and (e), the orange, violet and green lines describe, respectively, 
    the low-, intermediate- and high-filling situations considered in the text.}
  \label{spectra_SU2}
\end{figure}

In order to discuss the helical properties of this system, a good starting point is
the analysis of the non-interacting physics for the $\mathcal{I}=1/2$ case. 
The single-particle spectrum of the Hamiltonian $\hat{\mathcal H}$
has two branches with the following dispersion relations:
\begin{equation}
  \epsilon_\pm(k_p) = -2t \cos \frac \gamma2 \, \cos k_p 
  \pm \sqrt{4t^2 \sin^2 \frac \gamma2 \, \sin^2 k_p + \Omega^2} \, .
  \label{spectrum_formula}
\end{equation}
When the condition $\Omega  < 2t \sin \frac \gamma2 \tan \frac \gamma2$ is satisfied, 
the lower branch displays two minima at $k_p \approx \pm\gamma/2$ and a local maximum 
at $k_p=0$, see Fig.~\ref{spectra_SU2}(a): 
this case will be referred to as the weak-Raman-coupling (WRC) regime. 
In the opposite case, dubbed strong-Raman-coupling (SRC) regime, the lower branch 
has one single minimum at $k_p=0$ without any special feature at $k_p\neq0$, 
see Fig.~\ref{spectra_SU2}(c).

The study of the spin polarization $S^z$ (related to the operator
$\sum_{j,m} \hspace{-0.1cm} m \, \hat{\nu}_{j,m} \,$) 
of each eigenmode highlights an important difference 
between the SRC and the WRC regimes, see Figs.~\ref{spectra_SU2}(b) and~\ref{spectra_SU2}(d). 
In the WRC case, for most of the values of $k_p$, the eigenstates are prevalently polarized 
along the $z$ direction, while in the SRC regime this is not true 
(the dominating polarization is along the $x$ direction, not shown here). 
Figure~\ref{spectra_SU2}(a) also shows that in the WRC regime depending on the filling, 
the low-energy excitation may have very different properties.
For low (e.g. the orange line) or high (e.g. the green line) fillings,
there are four low-energy excitations. However, when the chemical potential 
(here we consider zero temperature) lies between $-2t \cos(\gamma/2)-\Omega$ 
and $-2t \cos (\gamma/2)+\Omega$ (e.g. the violet line),
there are two gapless excitations which have definite quasi-momentum 
and definite spin in the $z$ direction.
In the non-interacting case, this is an {\it helical liquid} which, 
once interpreted as a ladder, features two chiral edge modes.
 
Similar considerations about the single-particle spectrum hold for the $\mathcal{I}=1$ case, 
even though the analytic form of the eigenenergies is more involved. 
In Fig.~\ref{spectra_SU2}(e) we show the single-particle energy spectrum 
of the eigenstates in the WRC regime.
Low, intermediate and high fillings can be identified also in this case, and are 
indicated by the three different horizontal lines.
The intermediate filling (violet line) corresponds to the regime where the helical liquid appears; 
indeed the spin polarization $S^z$ shown in Fig.~\ref{spectra_SU2}(f)
exhibits almost full polarization of the eigenstates close to the considered Fermi energy.
Here, in the synthetic-dimension representation, the three-leg ladder displays chiral modes.

In the interacting case, the spectral properties of the Hamiltonian are 
not trivially computable. In the following section we define 
the physical quantities used to properly characterize the helical modes,
which can be calculated by means of the DMRG algorithm.
In the remainder of this paper we carefully analyze such quantities.

\section{Observables}
\label{obs-sec}

The study of the momentum distribution function, both spin-resolved 
and non-spin-resolved, can provide, as we shall see, information about 
the helical/chiral nature of the interacting liquid under consideration.
The spin-resolved momentum distribution function is defined as
\begin{equation}
  n_{p,m} = \langle \hat c^\dagger_{p,m} \hat c_{p,m}\rangle = 
  \frac{1}{L} \, \sum_{j,l} \, e^{-i\frac{2\pi p}{L}(j-l)} \langle\hat{c}_{j,m}^\dagger \hat{c}_{l,m} \rangle \,,
  \label{eq:n_k}
\end{equation}
where expectation values are taken over the ground state.
Since $p$ is not a good quantum number for $\hat{H}$, we will conveniently consider Hamiltonian $\hat{\mathcal{H}}$ and
the momentum distribution function $\nu_{p,m} = \langle \hat d^\dagger_{p,m} \hat d_{p,m} \rangle$, for which it easy to verify that 
$\nu_{p,m}=n_{p-m\gamma,m}$. 
Accordingly, the total momentum distribution is given by
$n_p = \sum_{m=-\mathcal{I}}^{\mathcal{I}} n_{p,m}$.

Based on these definitions, we introduce two chirality witnesses, i.e. two quantities 
which diagnose and identify the edge currents determined by the presence of the 
gauge field $\gamma \neq0$, even in the presence of repulsive interactions.
To this aim, we first solve the continuity equation for the Hamiltonian $\hat{{H}}$ 
and define the ground-state average chiral current 
\begin{equation}
  \mathcal{J}_{j,m} = -i \, t \, \langle  \hat{c}^\dagger_{j,m}  \hat{c}_{j+1,m} \rangle  + \text{H.c.}\;.
  \label{cur}
\end{equation}
Assuming PBC in the real dimension and using Eq.~(\ref{eq:n_k}), 
its spatial average can be re-expressed as
\begin{equation}
  Q_m = \frac{1}{L} \sum_j  \mathcal{J}_{j,m} = 
  - \frac{2t}{L}\sum_{p>0}\sin k_p\left(n_{p,m}-n_{-p,m}\right) \,,  \label{QMM}
\end{equation}
with $k_p = 2\pi p / L$. The latter relation allows to indirectly probe
the existence of chiral currents using a quantity, namely $n_{p,m}$,
which can be experimentally observed in the state-of-art laboratories 
using a band-mapping technique~\cite{Kohl_2005} followed by a Stern-Gerlach 
time-of-flight imaging~\cite{Mancini_2015, Stuhl_2015}. 
The quantity $Q_m$ is the first chirality witness to be employed in the following.
 
The second chirality witness is the quantity
\begin{equation}
  J_m = -\sum_{p>0}\left(n_{p,m}-n_{-p,m}\right) \,,  \label{JM}
\end{equation}
defined in Ref.~\cite{Mancini_2015}, which is more directly related 
to the asymmetry of the spin-resolved momentum distribution function. 
Both $J_m$ and $Q_m$ give information about the circulating currents and, 
as we shall see below, display the same qualitative behavior 
(they only differ for a cut-off at low wavelength).

\section{Results}
\label{results}

Equipped with the definitions given in the previous sections, we now discuss 
how atom-atom repulsive interactions affect the momentum distribution functions $n_p$ 
and $n_{p,m}$ and the chirality witnesses $Q_m$ and $J_m$.
The results for the non-interacting cases, here used as a reference, are computed by means of exact 
diagonalization, while for $U/t\neq0$ the DMRG algorithm is used~\cite{White_1992, Schollwock_2011}. 
We only address the ground-state properties, i.e. rigorously work at zero temperature.
In the finite-size sweeping procedure, up to 250 eigenstates of the reduced density matrix 
are kept, in order to achieve a truncation error of the order of $10^{-6}$ (in the worst cases) 
and a precision, for the computed correlations, at the fourth digit. 
The resulting inaccuracy is negligible on the scale of all the figures shown hereafter.

Unless differently specified, in the $\mathcal{I}=1/2$ case we consider $L=96$ 
and $\Omega/t=0.3$, while in the $\mathcal{I}=1$ case we set $L=48$ and $\Omega/t=0.1$ 
(the ratio $\Omega/t$ is chosen in order to be in the WRC regime); 
$\gamma=0.37\pi$ coincides with the experimental value of Ref.~\cite{Mancini_2015}. 

As shown in Figs.~\ref{spectra_SU2}(a) and~\ref{spectra_SU2}(e), in the non-interacting regime 
we can outline three inequivalent classes of fillings that we dub low, intermediate 
and high. In the specific, we consider $N/L=3/16$, $3/8$ and $7/12$ for $\mathcal{I}=1/2$, 
and $N/L=1/4$, $13/24$ and $5/6$ for $\mathcal{I}=1$ 
corresponding to the low-, intermediate-, and high-filling cases respectively.
OBC in the real dimension have been adopted.

\subsection{Momentum distribution functions}

\begin{figure}
  \centering
  \includegraphics[width=\linewidth]{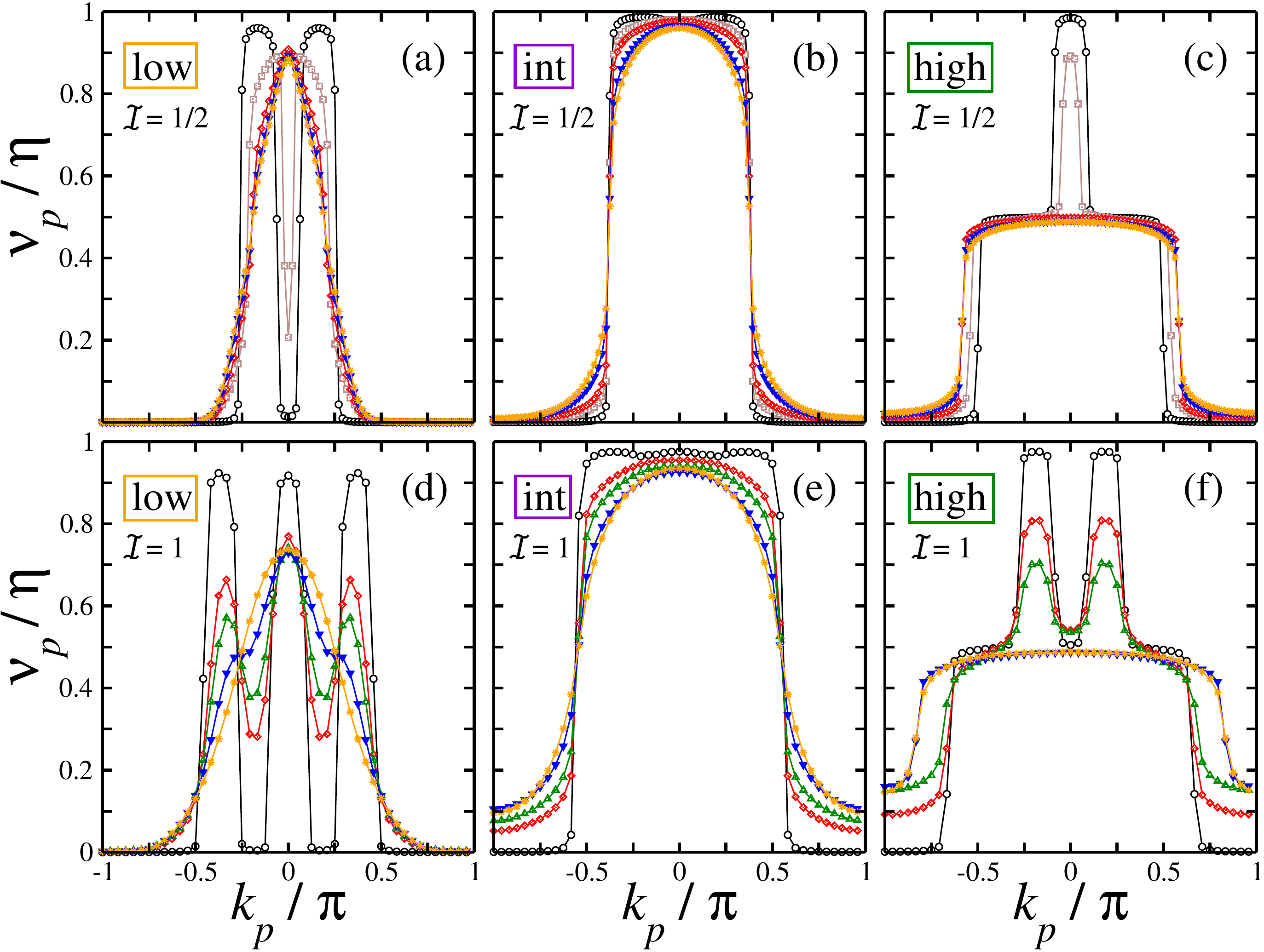}
  \caption{Momentum distribution functions $\nu_p$ for different values 
    of the interaction coefficient. 
    First row: $\mathcal{I}=1/2$; second row: $\mathcal{I}=1$. 
    First column: low-filling case ($\eta=1$); second column: intermediate-filling case ($\eta=1$); 
    last column: high-filling case ($\eta=2$). 
    The various colors denote different $U/t$ values: 
    0 (black circles), 3 (brown squares), 5 (red diamonds), 8 (green triangles up), 20 (blue triangles down), $ U/t\to \infty$ (orange stars).
    }
  \label{nkTOT}
\end{figure}

Let us first focus on the $\mathcal{I}=1/2$ case. 
In Figs.~\ref{nkTOT}(a-c) we plot the momentum distribution function $\nu_p$ 
for the three fillings listed above. For $U/t=0$, the behavior of $\nu_p$ 
can be easily predicted by looking at the single-particle spectrum: 
in the low and high-filling cases peaks arise in correspondence of the partially occupied 
energy wells, while in the intermediate-filling case a more homogeneous 
momentum distribution function emerges. 

The presence of repulsive atom-atom interactions significantly modifies 
the momentum distribution functions in the low- and high-filling cases: 
when $U/t$ is increased, they drive the distribution towards a more homogeneous shape 
with enhanced tails, a typical effect of interactions~\cite{Giamarchi_2003}. 
On the contrary, in the intermediate-filling case the homogeneous behavior is unmodified, 
apart from the mentioned tails. Such a phenomenology is well explained using bosonization 
and renormalization-group techniques, as discussed in Ref.~\cite{Braunecker_2010}.
Interactions lead to an effective enhancement of the energy of the two gapped modes, 
whose presence characterizes the helical liquid. 
Effectively, the interacting system behaves as if $\Omega/t$ were renormalized and increased, 
thus enhancing the filling regimes for which an helical liquid can be expected. 
Furthermore, this is in agreement with the fact that the non-interacting 
helical liquid is essentially left unchanged by the interactions. 
Thus, provided the interaction is sufficiently strong, even low- and high-filling setups 
can be driven into an helical liquid. This is the first important result of our analysis: 
repulsive interactions enhance the gap protecting the helical liquid.

The momentum distribution functions for $\mathcal{I}=1$ at the three cited fillings 
display the same qualitative behavior, see Figs.~\ref{nkTOT}(d-f). 
Again, the underlying physics can be explained in terms of an effective enhancement of $\Omega/t$, 
due to the presence of interactions. Contrary to the previous case, 
for values of $\mathcal I$ larger than $1/2$, no analytical prediction is available, 
but it seems reasonable to believe that a similar behavior should occur.

It is important to note that in the SRC regime on-site interactions are not expected 
to significantly modify the momentum distribution function of the non-interacting system. 
The occupied single-particle states belong only to the lowest band 
and are almost polarized in the same direction, $x$: the gas is thus quasi-spinless 
and an on-site interaction should only weakly alter the ground state because 
of Pauli exclusion principle. Additional numerical investigations may help in clarifying
this issue.

\begin{figure}
  \includegraphics[width=\linewidth,trim=0cm -0.3cm 0cm 0cm,clip=true]{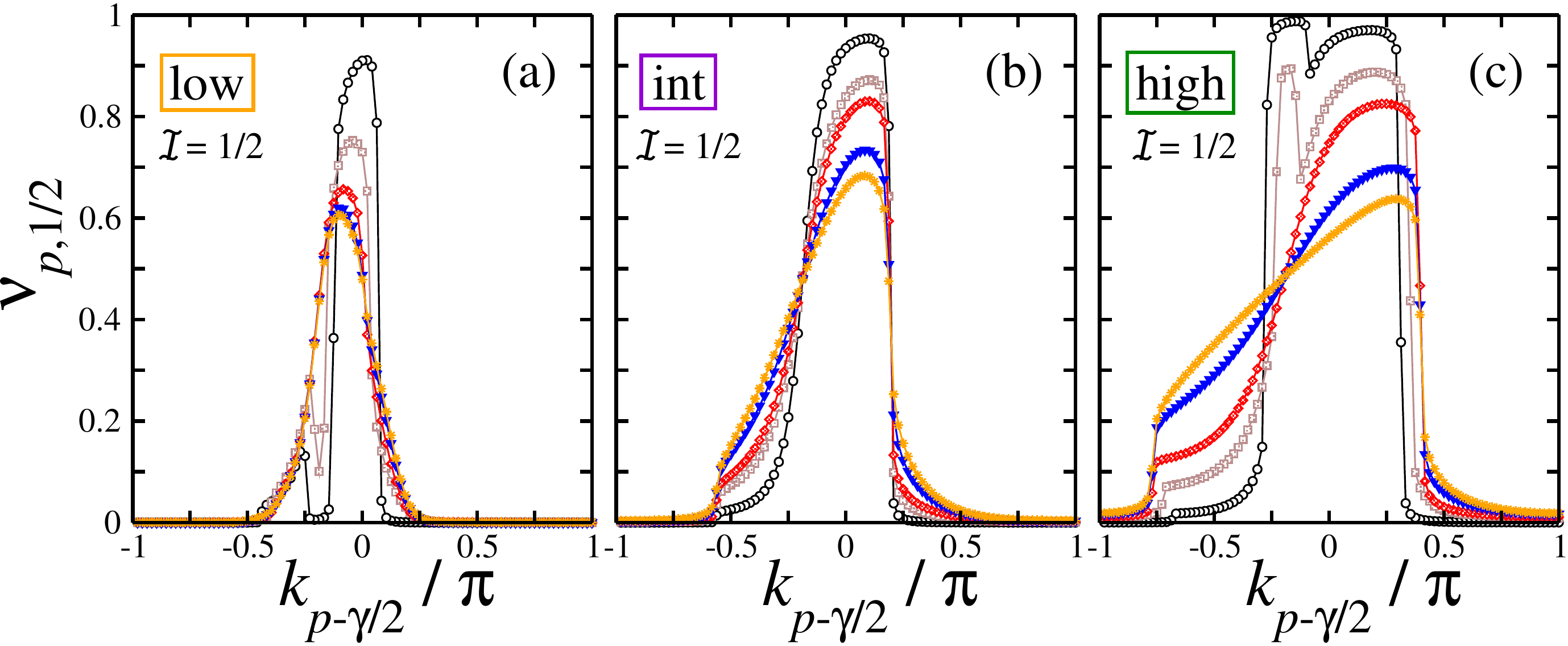}
  \includegraphics[width=\linewidth]{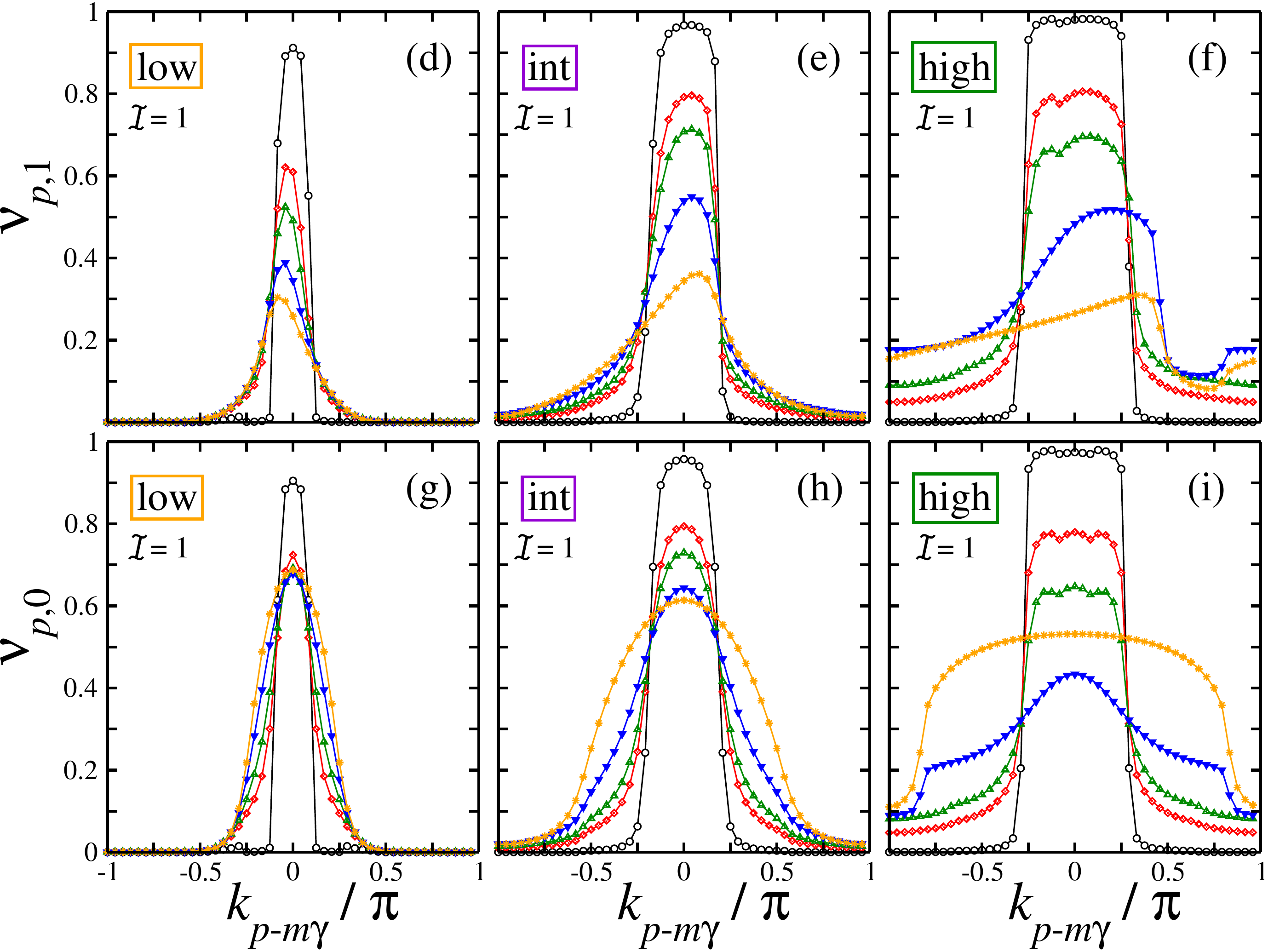}
  \caption{Spin-resolved momentum distribution functions $\nu_{p,m}$ 
    for different values of $U/t$ in the WRC regime. 
    First row: $\mathcal{I}=1/2$ (note that $\nu_{p,-1/2}=\nu_{-p,1/2}$);
    second and third row: $\mathcal{I}=1$ (note that $\nu_{p,-1}=\nu_{-p,1}$). 
    Panels (a), (d) and (g): low-filling case; panels (b), (e) and (h): intermediate-filling case; 
    panels (c), (f) and (i): high-filling case. 
    For the color code, see the caption of Fig.~\ref{nkTOT}.
    }
  \label{srnk}
\end{figure} 

Further information about the system can be revealed by the spin-resolved momentum 
distribution functions $\nu_{p,m}$.
In Figs.~\ref{srnk}(a-c) we plot such functions in the WRC regime 
for the spin species $m=1/2$ and $\mathcal I=1/2$. 
Such profiles are clearly asymmetric with respect to $k_p=0$, indicating the helical nature 
of the ground state. Note that the asymmetry is enhanced by the interactions. 
A similar behavior is observed for $m = \pm 1$ and $\mathcal I=1$, see Figs.~\ref{srnk}(d-f). 
On the other hand, for symmetry reasons, the momentum distribution function $\nu_{p,m=0}$ 
is symmetric with respect to $k_p=0$, although it is modified 
by the interactions, see Figs.~\ref{srnk}(g-i).

\subsection{Chirality witnesses}
\label{currents-sec}

\begin{figure}
  \centering
  \includegraphics[width=\linewidth]{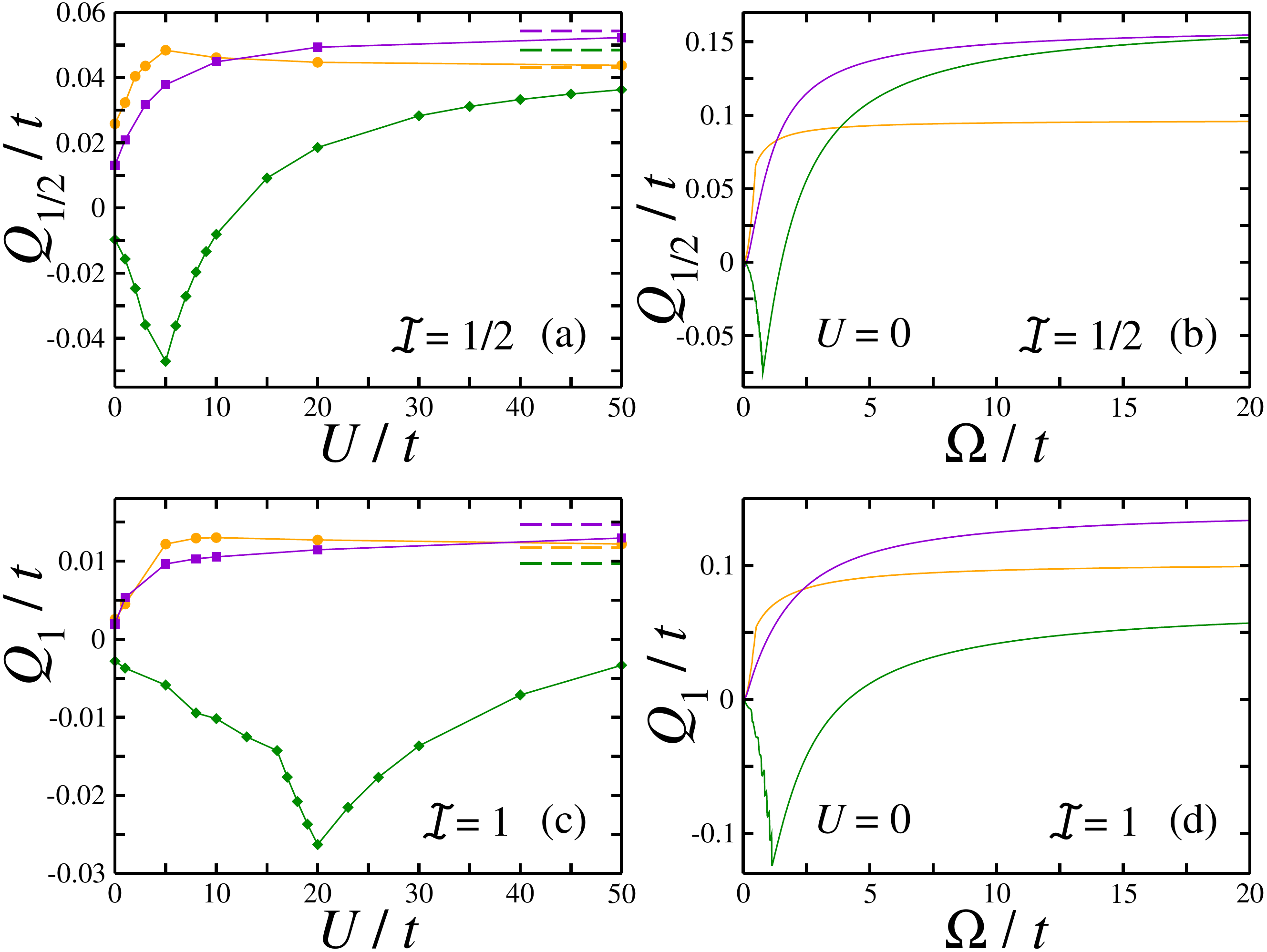}
  \caption{Dependence of $Q_{m=\mathcal I}$ on the interaction strength.
    Panel (a): $Q_{1/2}$ for $\mathcal{I}=1/2$ as a function of the interaction strength $U/t$; 
    dashed lines are the values of $Q_{1/2}$ in the limit $U/t\rightarrow\infty$. 
    Panel (b): $Q_{1/2}$ for $\mathcal{I}=1/2$ in the non-interacting case 
    ($U/t=0$) for different values of $\Omega/t$.
    Panels (c) and (d): same analysis for $\mathcal{I}=1$ and $m=1$.
    The various curves denote the different regimes of low (orange circles),
    intermediate (violet squares) and high (green diamonds) filling.}
  \label{Q-fig}
\end{figure}

In this paragraph we discuss the properties of the chirality witnesses $Q_m$ and $J_m$ 
for an interacting system. Even though a preliminary analysis of these quantities 
has been carried out in Ref.~\cite{Barbarino_2015}, a systematic study of the effects of repulsive 
atom-atom interactions in a relevant experimental setup~\cite{Mancini_2015} is still lacking.

In Figs.~\ref{Q-fig}(a) and~\ref{Q-fig}(c) we display the behavior of $Q_{m=\mathcal I}$ 
as a function of $U$ for the cases $\mathcal{I}=1/2$ and $\mathcal I=1$; 
we focus again on the three fillings outlined above. 
In~\ref{app:currents} we show that, although the system has OBC and
it is not homogeneous, averaging over many lattice sites yields a value
related to the bulk current.
A first striking observation is that one can observe different trends, 
also displaying non-monotonic features.
The role of interactions in protecting the helical liquid here encounters
a first naive confirmation:
in all cases, the value of $|Q_m|$ in the $U/t \to \infty$ limit exceeds 
that of the non-interacting system.

\begin{figure}
  \centering
  \includegraphics[width=\linewidth]{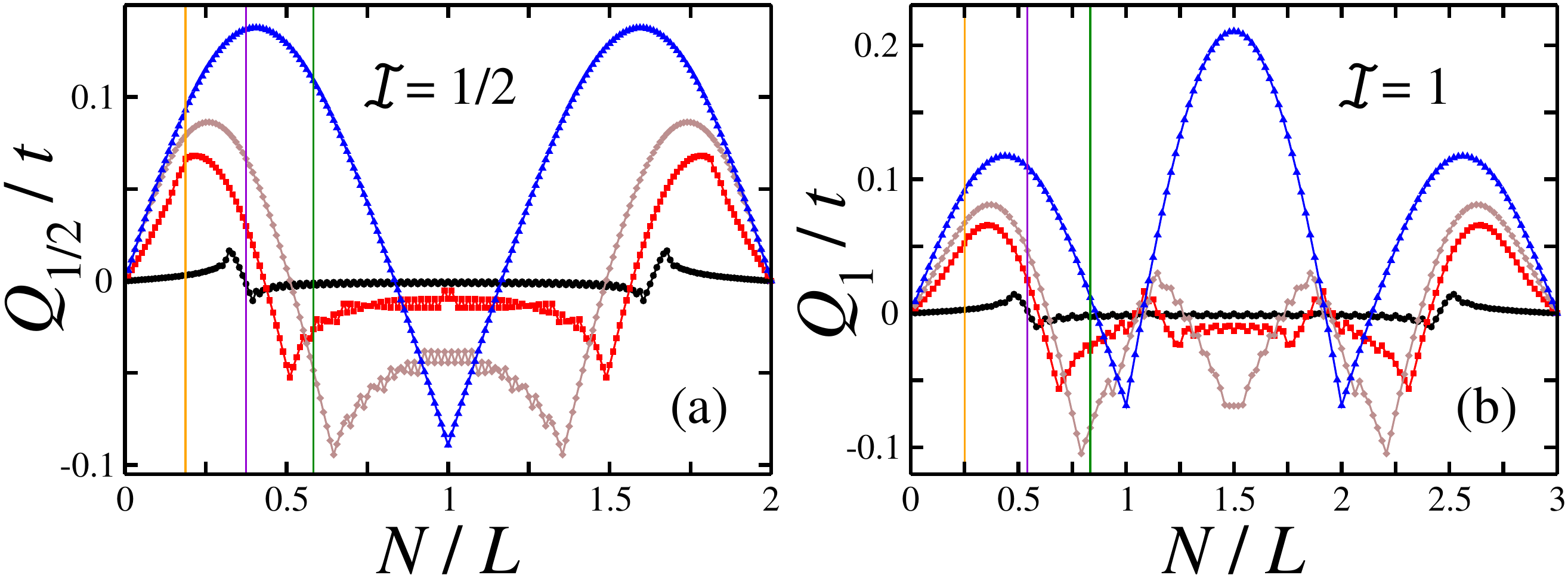}
  \caption{Spatially-averaged currents as a function of the density of atoms.
    Panel (a): $Q_{1/2}$ for $\mathcal{I}=1/2$ in the non-interacting case 
    and for different values of $\Omega$ 
    (black: $\Omega=0.1$, red: $\Omega=0.5$, brown: $\Omega=1$, blue: $\Omega=5$); 
    vertical lines mark low, intermediate and high fillings, with the same color code 
    as in Fig.~\ref{spectra_SU2}(a). 
    Panel (b): same analysis for $\mathcal{I}=1$ and $m=1$.}
  \label{Q-fig-nint}
\end{figure}

In order to understand the dependence of $Q_m$ on $U/t$, we employ an effective model.
We have already noticed that the most prominent effect of the interactions on $\nu_p$ 
is that of letting the system behave as if it were non-interacting but with 
a renormalized value of $\Omega$. Here we test this observation by studying the dependence 
of $Q_m$ on $\Omega$ in the absence of interactions. Results displayed 
in Figs.~\ref{Q-fig}(b) and~\ref{Q-fig}(d) show that this simple model 
offers a good qualitative understanding of the interacting system.
For example, in both the $\mathcal{I}=1/2$ and $\mathcal{I}=1$ cases, 
$Q_{m=\mathcal I}$ displays the same (quasi-)monotonic increasing behavior with $U/t$
and with $\Omega/t$, for the low and intermediate fillings. 
In the high-filling case, $Q_{m=\mathcal I}$ exhibits a strongly non-monotonic behavior 
as a function of $U$; in particular the plot points out a change in sign 
which is \textit{a priori} unexpected because in the classical case the 
magnetic field determines unambiguously the direction of the circulating currents.
To further elucidate this problem, in Fig.~\ref{Q-fig-nint} we plot the dependence
of $Q_m$ on the filling $N/L$ for a fixed value of $\Omega/t$ and $U/t=0$. 
The plot shows that at low fillings the value of $Q_{m=\mathcal I}$ increases gently, 
but experiences an abrupt decrease once the helical region is entered, 
marked by the violet line (intermediate fillings). 
For higher fillings (even outside the helical region) and for small $\Omega$, 
the value of $Q_{m=\mathcal{I}}$ is negative and thus the current changes sign; 
however, by increasing $\Omega$, $Q_{m=\mathcal{I}}$ also increases, 
crossing $0$ and becoming positive and finite.
It thus follows that in this system the chiral currents 
are not strictly speaking chiral and  states with opposite current flow occur at accessible energies.

\begin{figure}
  \centering
  \includegraphics[width=\linewidth]{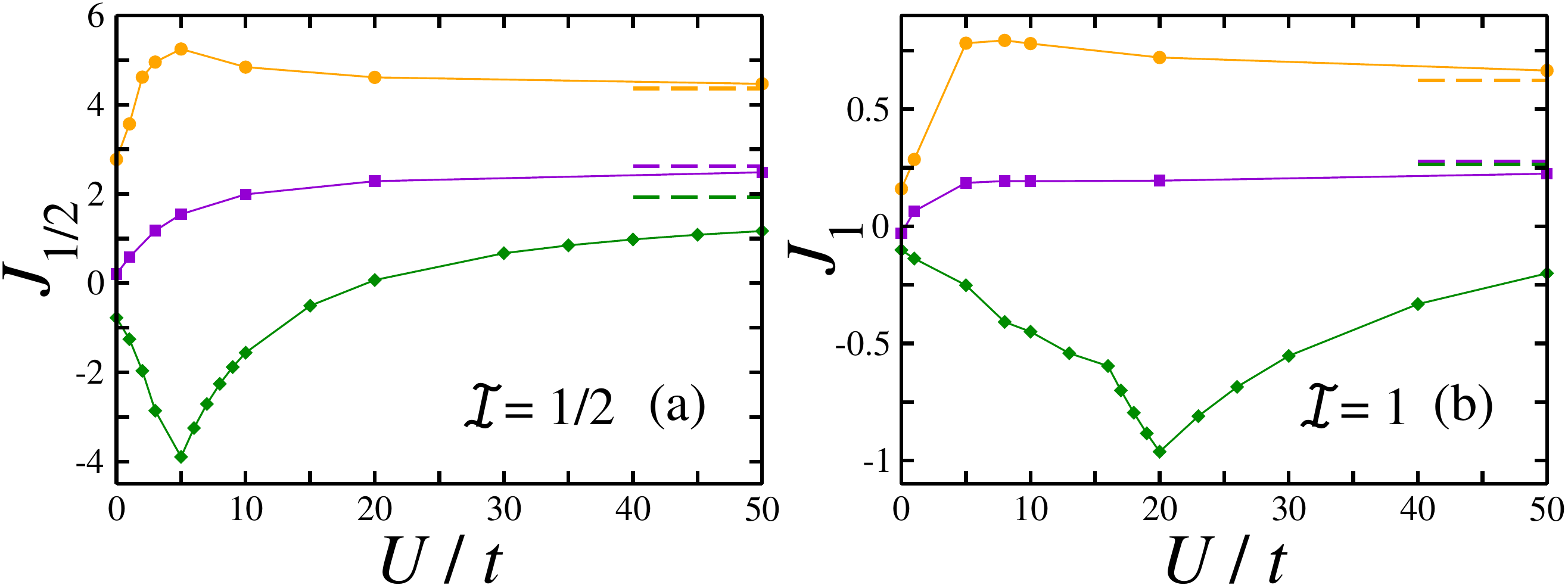}
  \caption{Dependence of $J_{m=\mathcal I}$ on the interaction strength.
    Panel (a): $Q_{1/2}$ for $\mathcal{I}=1/2$ at low (orange circles), intermediate (violet squares) and high filling (green diamonds) as a function of the interaction strength $U/t$; 
    dashed lines denote the values of $J_{1/2}$ in the limit $U/t\rightarrow\infty$. 
    Panel (b): same analysis for $\mathcal{I}=1$ and $m=1$.}
  \label{J-fig}
\end{figure}

The chirality witness $J_{m=\mathcal I}$ shares many similarities with $Q_{m=\mathcal I}$. 
In Fig.~\ref{J-fig} we plot $J_{m = \mathcal I}$ as a function of $U$,
to be compared with Figs.~\ref{Q-fig}(a) and~\ref{Q-fig}(c) for $Q_{m=\mathcal I}$. 
Again, in the low- and intermediate-filling regimes $J_{m = \mathcal I}$ is almost monotonous, 
whereas monotonicity is significantly broken for high fillings. 
The explanation of this behavior can again be sought in the peculiar dependence 
of the current carried by the eigenmodes of the system.

\section{Conclusions}
\label{conclusions}

By means of DMRG simulations, we have studied the impact of atom-atom repulsive interactions 
on the quantum-Hall-like chiral currents recently detected in Refs.~\cite{Mancini_2015, Stuhl_2015}. 
We have modeled the experimental setup of Ref.~\cite{Mancini_2015} and characterized
the behavior of the edge currents through the asymmetry of the momentum distribution function.

We have considered different particle fillings and identified the filling range
where a chiral/helical liquid appears (in the text dubbed as ``intermediate'').
When the filling is slightly higher or lower, in the presence of repulsive interactions,
the system starts behaving as the non-interacting chiral/helical liquid. 
This leads to the first conclusion that interactions stabilize such phase.
To better assess its nature, we have introduced two chirality witnesses, which are displayed in
Figs.~\ref{Q-fig} and~\ref{J-fig}, where the chirality of the currents 
is studied as a function of the interaction strength $U/t$. 
As highlighted in the plots, the role of the interaction is non-trivial,
and in the strongly-repulsive limit leads to the enhancement of the persistent currents.

In the analysis presented here we have neglected the role of an harmonic trapping
confinement as well as finite-temperature effects. Their interplay with interactions
and the edge physics highlighted so far is left for a future work.

The edge currents studied here do not have a topological origin. However, these 
synthetic ladders may support fractional quantum Hall-like states~\cite{Barbarino_2015, Cornfeld_2015}, and it would be very interesting to understand how to explore this regime by means of the quantities 
discussed in the present paper. In particular it would be important to develop a complete characterization of how fractional quantization may emerge in a cold atomic setup.

\section*{Acknowledgements}
We thank Leonardo Fallani and Guido Pagano for enlightening discussions, and S. Sinigardi for technical support. 
We acknowledge INFN-CNAF for providing us computational resources and support, and D. Cesini in particular. 
We acknowledge financial support from the EU integrated projects SIQS and QUIC, from Italian MIUR 
via PRIN Project 2010LLKJBX and FIRB project RBFR12NLNA. R.F acknowledges the Oxford Martin School 
for support and the Clarendon Laboratory for hospitality during the completion of the work. 

\appendix

\section{Currents}
\label{app:currents}

The chirality witness $Q_m$ is the space-average value of the expectation value 
of the current operator over the ground state of the system, $\mathcal J_{j,m}$. 
Whereas in a homogeneous system with PBC this value coincides with the expectation value 
of the current on every site, the effects of the boundaries in a system with OBC 
might play an important role.

\begin{figure}
  \centering
  \includegraphics[width=\linewidth]{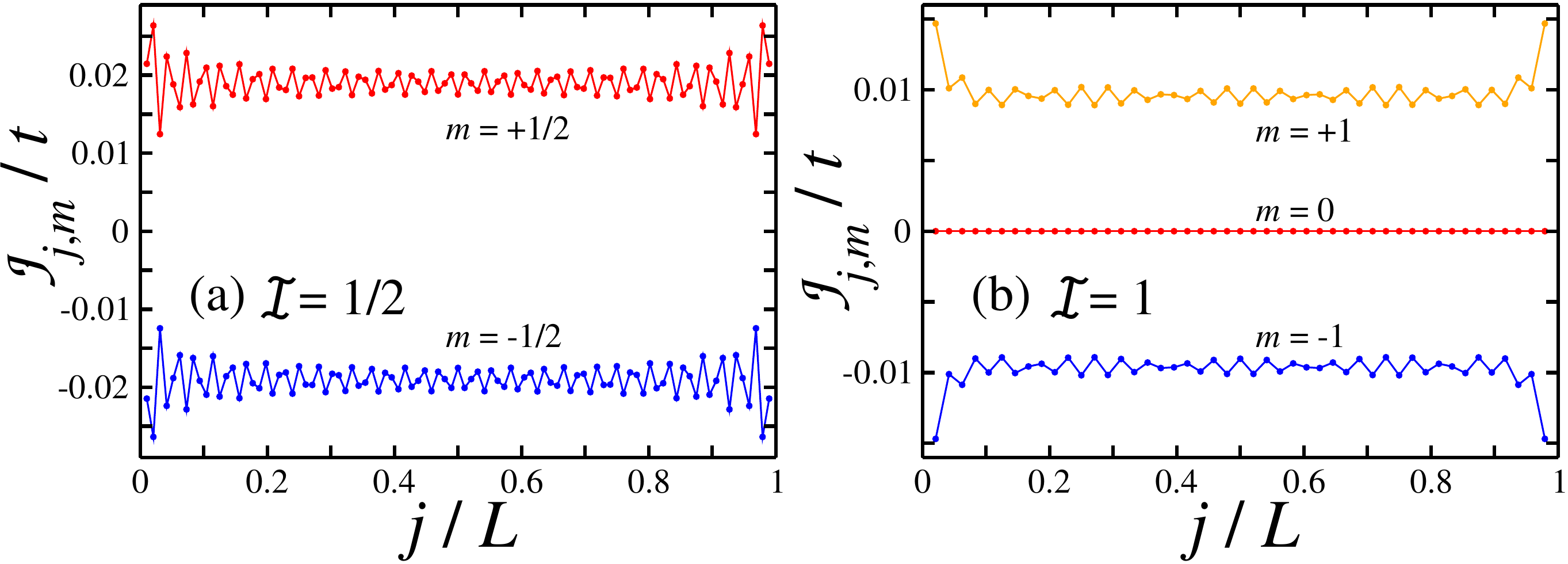}
  \caption{Spatial profile of the spin-resolved currents $\mathcal J_{j,m}$. 
    Panel (a): $\mathcal{I}=1/2$ (blue: $m=-1/2$; red: $m=1/2$). 
    Panel (b): $\mathcal{I}=1$ (blue: $m=-1$; red: $m=0$; orange: $m=1$). 
    In both cases, intermediate filling and $U/t=5$ were chosen. 
    The color code refers to Fig.~\ref{ladder}. 
    The other parameters of the simulations are set as in Sec.~\ref{results}.
    }
  \label{cur-fig}
\end{figure}

In Fig.~\ref{cur-fig} we plot $\mathcal J_{j,m}$ both for a system with 
$\mathcal I=1/2$ [panel (a)] and with $\mathcal{I}=1$ [panel (b)].
The important information contained in the figure is that even if the system 
is clearly inhomogeneous, the space pattern of $\mathcal J_{j,m}$ is that of a small 
and fast oscillation over a constant value, so that the space average is an indicative 
quantity of the underlying physics. 
For both $\mathcal{I}=1/2$ and $\mathcal{I}=1$ the oscillations vanish 
in the limit $L\rightarrow +\infty$, see Ref.~\cite{Barbarino_2015}.

\section*{References}

\bibliographystyle{iopart-num}

\end{document}